\documentclass[%
reprint,
 amsmath,amssymb,
 aip,
 rsi,
]{revtex4-2}

\usepackage{gensymb,units,textcomp,color,leftidx}
\usepackage[caption=false]{subfig}
\usepackage[acronym]{glossaries}
\usepackage{graphicx}
\usepackage{bm}
\usepackage{xspace}
\usepackage{hyperref}

\usepackage{lipsum}

\newcommand{\CELIA}{Université de Bordeaux-CNRS-CEA, Centre Lasers Intenses et Applications (CELIA),\\ UMR 5107, F-33405 Talence, France}
\newcommand{\UVa}{Departamento de F\'isica Te\'orica At\'omica y \'Optica, Universidad de Valladolid, 47011 Valladolid, Spain}
\newcommand{\LLNL}{Lawrence Livermore National Laboratory, Livermore, California 94550, USA}
\newcommand{\ULPGC}{iUNAT–Departamento de F\'isica, Universidad de Las Palmas de Gran Canaria, 35017 Las Palmas de Gran Canaria, Spain}
\newcommand{\GA}{General Atomics, San Diego, California 92121, USA.}
\newcommand{\ICL}{Blackett Laboratory, Imperial College London, London, SW7 2AZ, UK}
\newcommand{\UCSD}{Center for Energy Research, University of California-San Diego, La Jolla, California 92093, USA}

\newacronym{xrfc}{XRFC}{X-Ray Framing Cameras}
\newcommand{\xrfc}{\gls{xrfc}\xspace}
\newacronym{fwhm}{FWHM}{Full Width at Half Maximum}

\newacronym{MIFEDS}{MIFEDS}{magneto-inertial fusion electrical discharge system}
\newcommand{\MIFEDS}{\gls{MIFEDS}\xspace}
\newacronym{MHD}{MHD}{magnetohydrodynamic}
\newcommand{\MHD}{\gls{MHD}\xspace}
\newacronym{HED}{HED}{High Energy Density}
\newcommand{\HED}{\gls{HED}\xspace}
\newacronym{ICF}{ICF}{Inertial Confinement Fusion}
\newcommand{\ICF}{\gls{ICF}\xspace}
\newacronym{MagLIF}{MagLIF}{Magnetized Liner Inertial Fusion}
\newcommand{\maglif}{\gls{MagLIF}\xspace}

\newcommand{\los}{line of sight\xspace}

\begin{document}


\title{X-ray imaging and radiation transport effects on cylindrical implosions}

\author{G. P\'{e}rez-Callejo}
	\email{gabriel.perez.callejo@uva.es}
	\affiliation{\UVa}
\author{M. Bailly-Grandvaux}
    \affiliation{\UCSD}
\author{R. Florido}
    \affiliation{\ULPGC}
\author{C. A. Walsh}
    \affiliation{\LLNL}
\author{M. A. Gigosos}
    \affiliation{\UVa}
\author{F. N. Beg}
    \affiliation{\UCSD}
\author{C. McGuffey}
    \affiliation{\GA}
\author{R. C. Mancini}
    \affiliation{Physics Department, University of Nevada, Reno, Nevada 89557, USA}
\author{F. Suzuki-Vidal}
    \affiliation{\ICL}
\author{C. Vlachos}
    \affiliation{\CELIA}
\author{P. Bradford}
    \affiliation{\CELIA}
\author{J. J. Santos}
    \affiliation{\CELIA}

\begin{abstract}

Magnetization of inertial confinement implosions is a promising means of improving their performance, owing to the potential reduction of energy losses within the target and mitigation of hydrodynamic instabilities. In particular, cylindrical implosions are useful for studying the influence of a magnetic field thanks to their axial symmetry. Here we present experimental results from cylindrical implosions on the OMEGA-60 laser using a 40-beam, \unit[14.5]{kJ}, \unit[1.5]{ns} drive and an initial seed magnetic field of $B_0=\unit[24]{T}$ along the axis of the targets, compared with reference results without an imposed B-field. Implosions were characterized using time-resolved X-ray imaging from two orthogonal lines of sight. We found that the data agree well with magnetohydrodynamic simulations once radiation transport within the imploding plasma is considered. We show that for a correct interpretation of the data in this type of experiments, explicit radiation transport must be taken into account.
\end{abstract}

\maketitle

\section{Introduction}

The effect of an external magnetic field (B-field) on \ICF implosions \cite{perkins2017} is a topic of ongoing interest in the \maglif \cite{gomez2014PRL}, indirect \cite{Moody_2020} and direct \cite{Gotchev_2009} drive communities. In laser-driven \ICF, seed  magnetic fields amplified by magnetic flux conservation during the implosion have the potential to increase fusion yields by relaxing the areal density requirement for ignition. In particular, cylindrical implosions are useful for studying these effects, as the B-field can be applied along the axis of the targets. The B-field compressed within the target acts in addition to inertia to confine the hot spot, resulting in a hotter fuel \cite{walsh2019pop}. This opens up the possibility of high-gain implosions with lower convergence ratios that are less susceptible to hydrodynamic instabilities. Magnetic fields can also effectively confine D-T ions and thermonuclear $\alpha$-particles \cite{sio2021}, enhancing collisionality and fusion yield \cite{hansen2020}.

The interpretation of magnetized implosion experiments relies heavily on comparisons with \MHD codes. These codes must account for extended-\MHD effects to accurately model energy and magnetic flux transport mechanisms within the plasma\cite{walsh2020pop}. To add confidence on their modelling capacity of more complicated scenarios of magnetized high-energy-density plasmas, the underlying physics requires to be benchmarked against experimental measurements in a simplified geometry and a priori easy-to-interpret regime. Characterizing the evolution of a cylindrical implosion and the compression of the fuel is fundamental to this benchmarking process (see Palaniyappan \textit{et al.}\cite{palaniyappan2020}, Sauppe \textit{et al.} \cite{sauppe2020} and references therein).

\begin{figure}
    \centering
    \includegraphics[width=0.5\columnwidth]{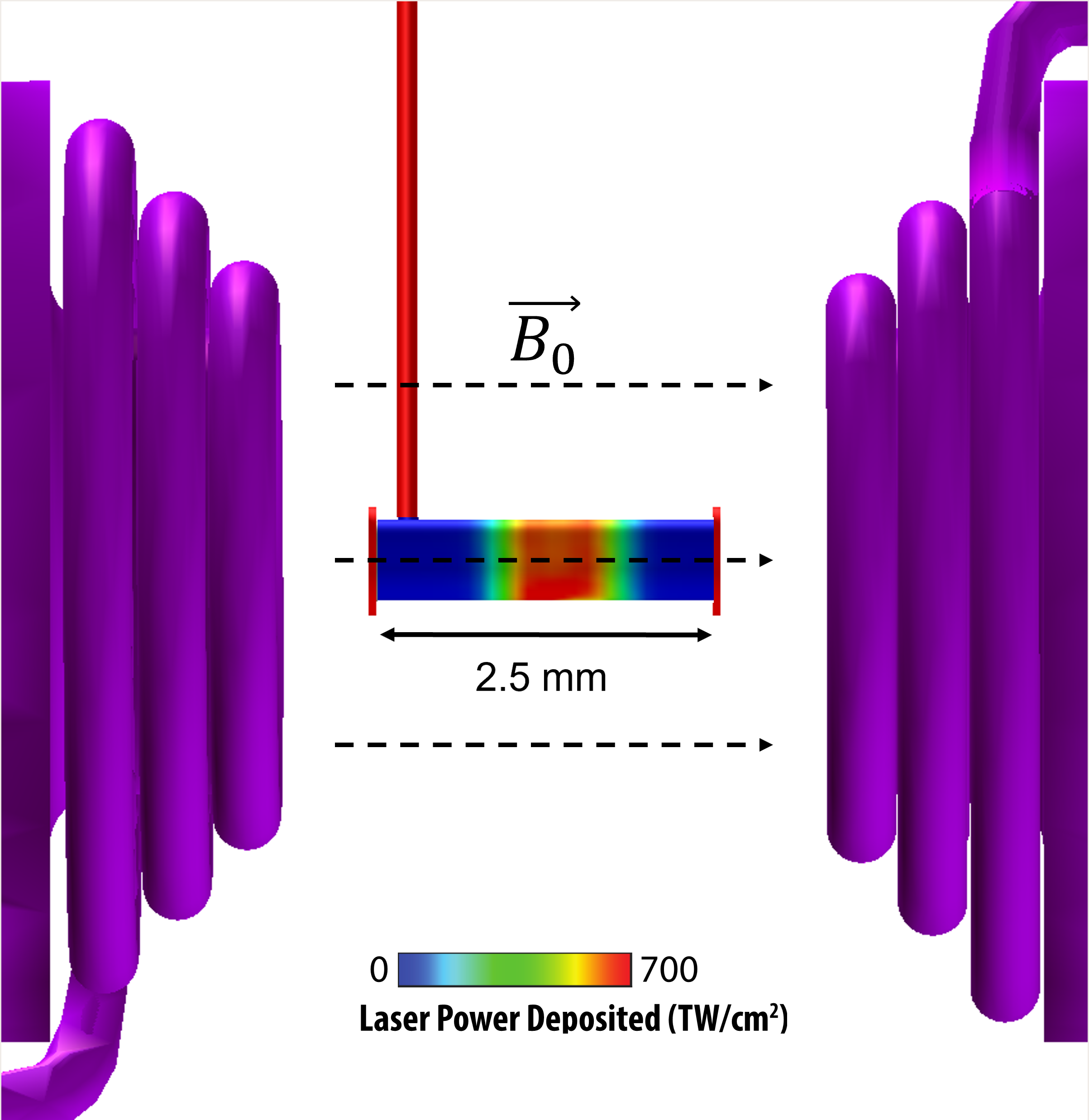}
    \caption{Overview of the experimental set up. The red stalk connected to the gas cylinder corresponds to the target holder and gas-fill, whereas the purple coils indicate the position in which the MIFEDS is placed in the shots. The direction of the seed B-field ($B_0=\unit[24]{T}$) is shown schematically. The colorscale on the cylinder corresponds to the laser irradiation profile. When the MIFEDS is fielded, the axial \los (along the cylinder axis) is blocked.}
    \label{fig:setup}
\end{figure}


\begin{figure*}
\centering
\subfloat[\label{fig:XRFC_Data}]{%
   \includegraphics[width=0.75\columnwidth]{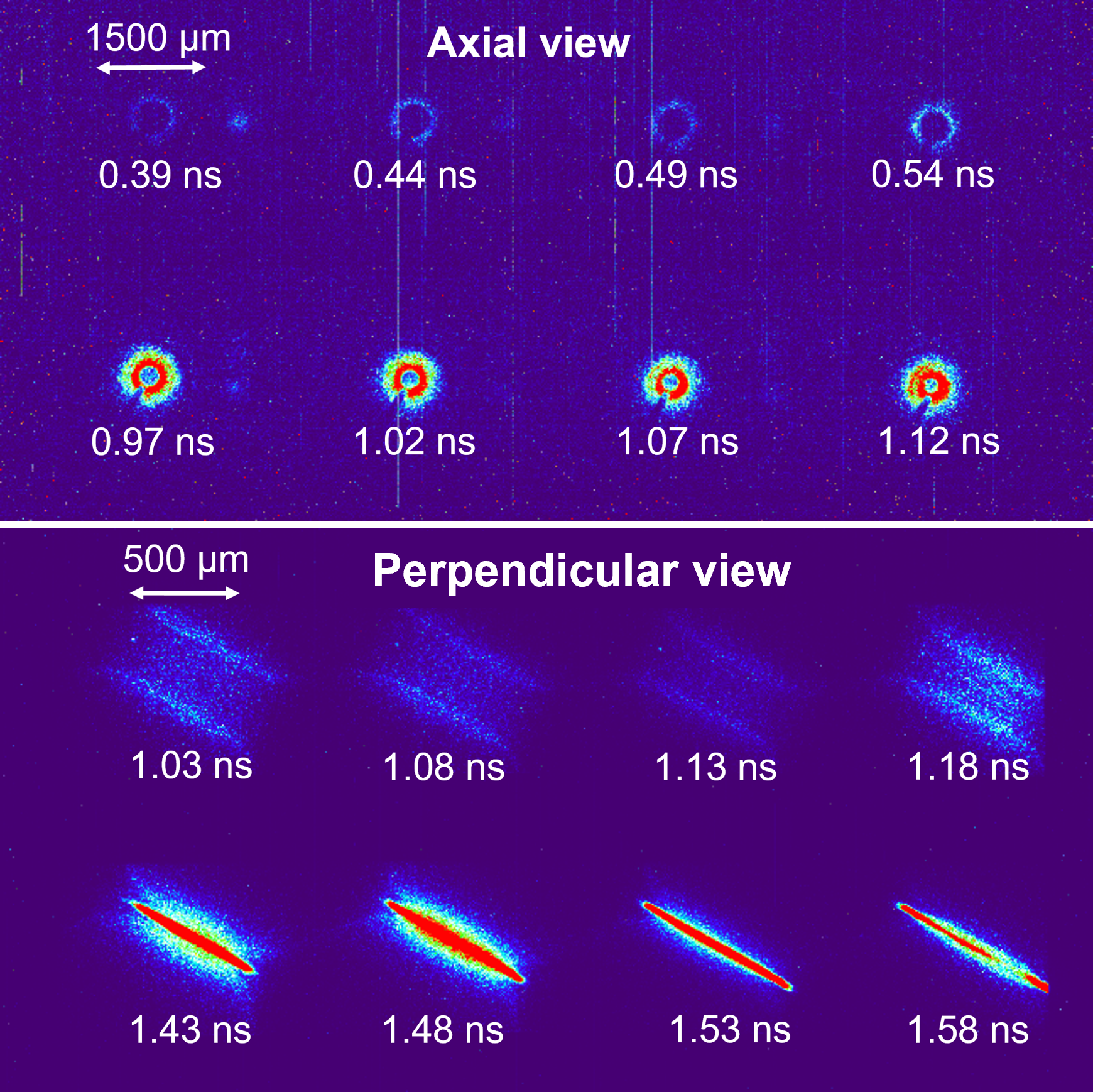}
  }
\subfloat[\label{fig:XRFC_Synthetic}]{%
    \includegraphics[width=0.75\columnwidth]{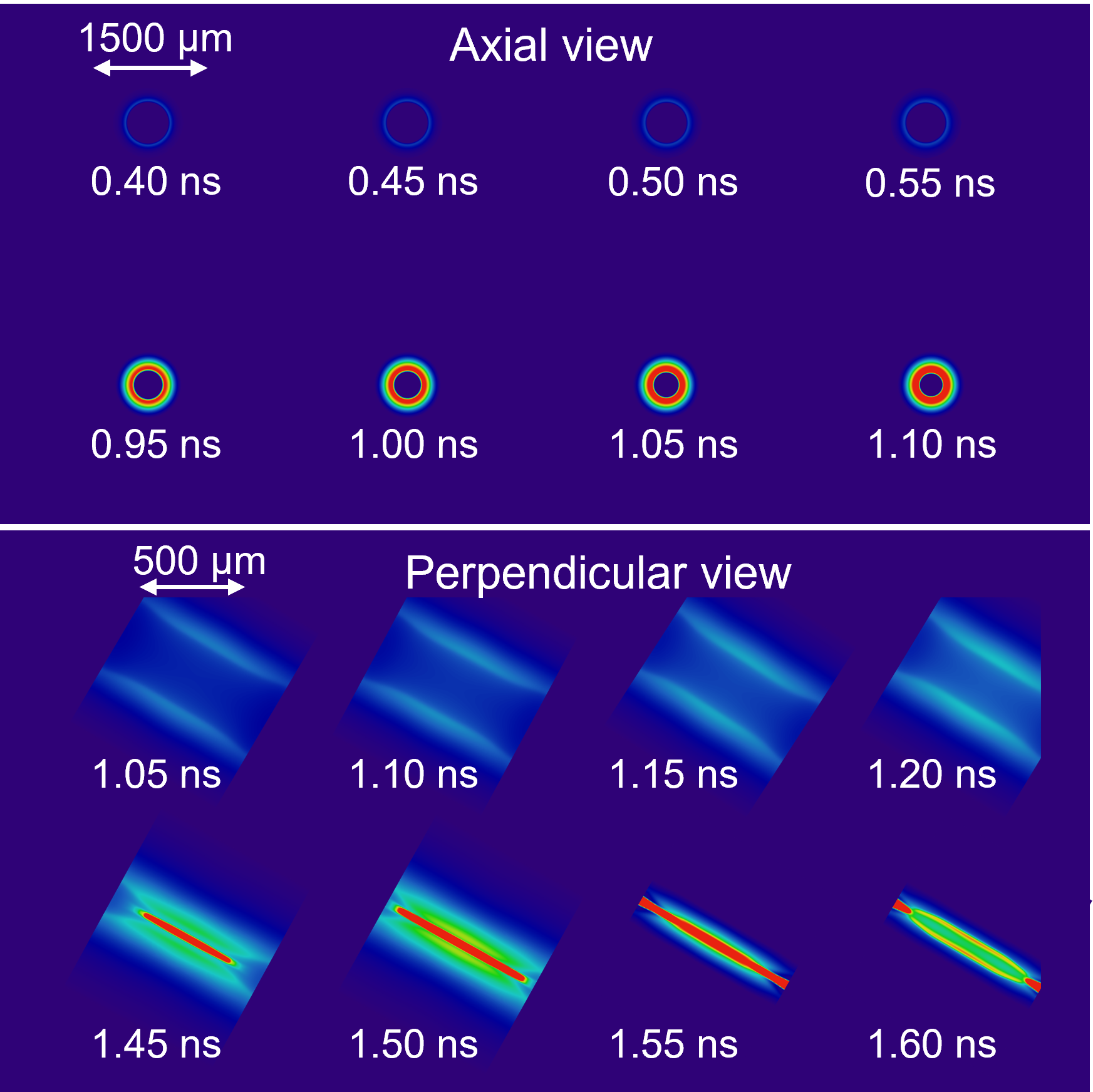}
}
\caption{(\ref{fig:XRFC_Data}) Example data of cylindrical implosions from XRFCs. The top part of the figure corresponds to 8 frames from the axial camera, whereas the bottom part shows 8 frames from the XRFC with a view normal to the cylinder axis. Note that the magnification (and therefore the scale) is different for each view. (\ref{fig:XRFC_Synthetic}) Corresponding synthetic 2D images produced by postprocessing the Gorgon simulations applying radiation transport and correcting for the instrument response.}
\label{fig:XRFC_Data_Image}
\end{figure*}


In this work, we present X-ray imaging data from experiments with laser-driven, magnetized cylindrical implosions similar to the mini-MagLIF concept explored at the OMEGA-60 laser \cite{davies2017}. We used two orthogonal \xrfc to record an axial and a perpendicular view of the cylinder, mapping the whole implosion up to the point of stagnation. We found that, as an effect of radiation transport within the imploding plasma, the apparent position of the shell is systematically shifted from its real value. This platform is a simple testbed for exploring magnetized phenomena in \HED plasmas, and the results presented here are a first step towards validating theoretical studies of this scenario \cite{walsh2022}. The paper is structured as follows: in Sec. \ref{sec:setup} we describe the experimental setup, physical parameters and the imaging cameras that were used, together with the details of the simulations performed with the Gorgon \MHD code \cite{ciardi2007,chittenden2004,walsh2017}. In Sec. \ref{sec:results} we summarise the experimental results and compare them with postprocessed simulations. Finally, our conclusions are presented in Sec. \ref{section:Conclusions}.

\section{Experimental set-up and modelling}
\label{sec:setup}

\begin{figure*}
\centering
\subfloat[\label{fig:XRFC_shell}]{%
   \includegraphics[width=0.8\columnwidth]{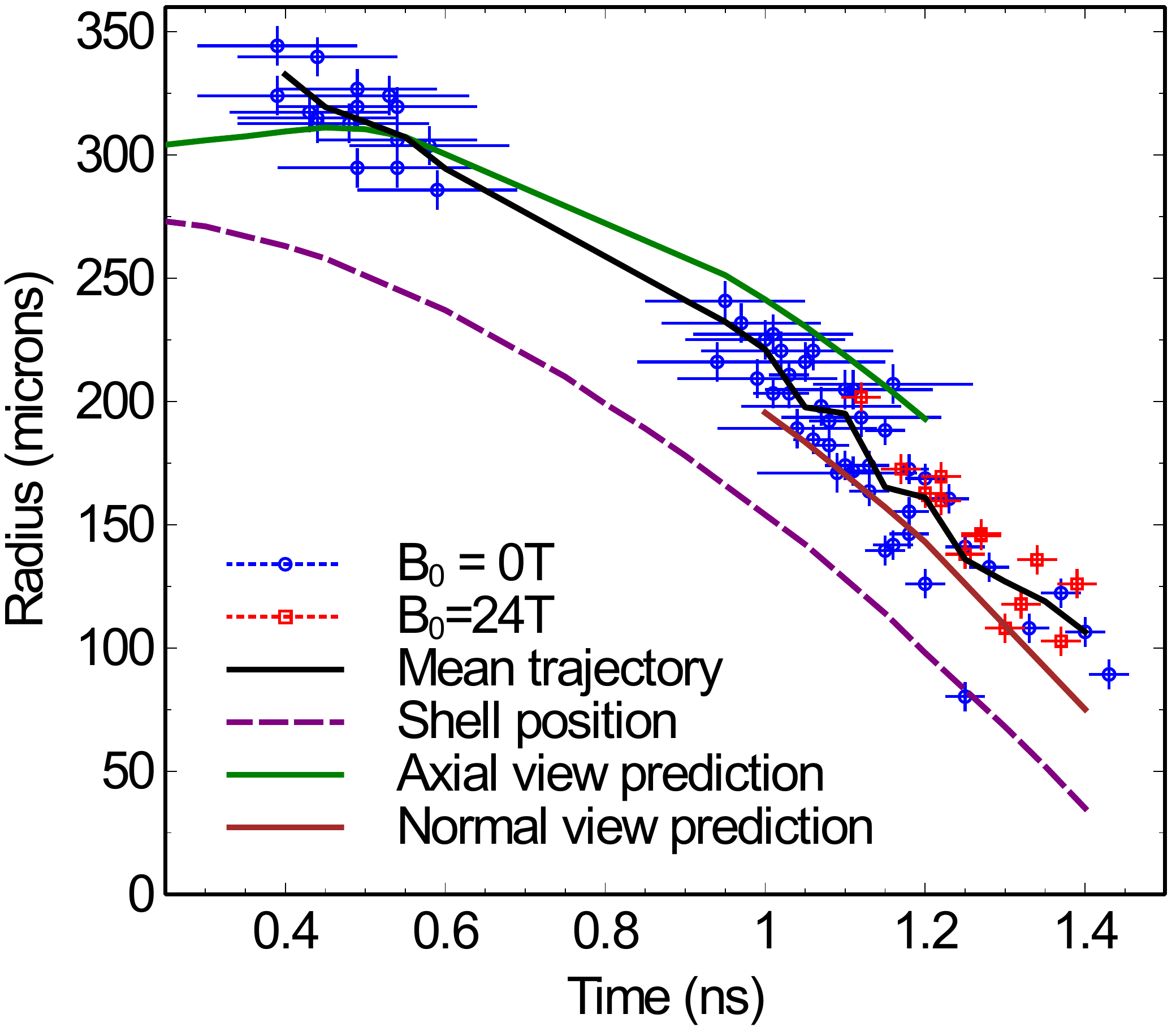}
  }
\subfloat[\label{fig:XRFC_core}]{%
    \includegraphics[width=0.8\columnwidth]{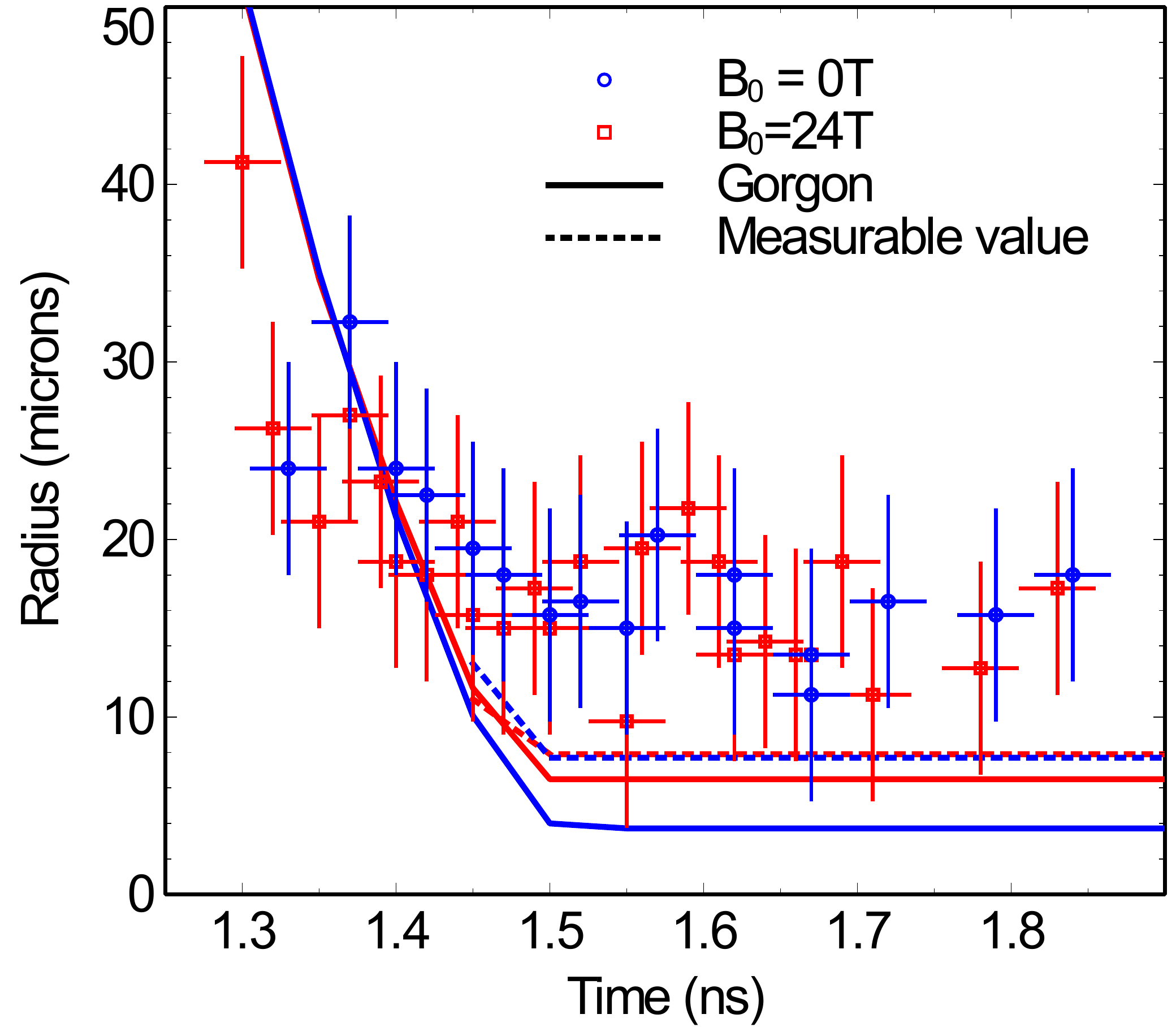}
}
\caption{\label{fig:XRFC_Plot} Radius of the implosion measured from the (a) separation of the shell walls and (b) width of the core. The blue circles and red squares correspond to implosions without B-field, and with a seed B-field of \unit[24]{T} respectively. Data were obtained from 9 shots with and without a seed B-field. In Fig. \ref{fig:XRFC_shell}, the purple dashed line indicates the trajectory of the shell obtained from the Gorgon simulations, whereas the green and brown lines correspond to the \textit{apparent} trajectory after postprocessing the simulations . The simulations do not predict any differences in the implosion owing to the B-field for the first \unit[1.4]{ns}. In Fig. \ref{fig:XRFC_core}, the solid lines correspond to the Gorgon prediction up to stagnation (prior to postprocessing) for the compressed radius (the color of the lines indicates the seed B-field consistently). The dotted lines correspond to the results from Gorgon after accounting for radiation transport and instrument response. Note how the differences in core compression caused by the B-field are washed out by radiation transport and could not be experimentally distinguished.}
\end{figure*}

The experiments (Fig. \ref{fig:setup}) were conducted on the OMEGA-60 laser, using a 40-beam, \unit[1.5]{ns}, \unit[14.5]{kJ}, 3$\omega$ laser drive to implode gas-filled cylindrical targets. The targets were \unit[2.5]{mm}-long Parylene-N tubes with an outer radius of $\unit[296\pm3]{\micro m}$ and a shell thickness of $\unit[18.2\pm1.3]{\micro m}$. The cylinders were filled with D$_2$ gas at \unit[11]{atm} ($\rho = \unit[1.81]{mg~cm^{-3}}$) and their pressure was monitored through a transducer connected to the target stalk on the target holder. An argon dopant (atomic concentration of 0.15\%) was added to the fuel as a spectroscopic tracer to infer the conditions of the compressed core at stagnation. The targets and laser drive were in line with previous mini-MagLIF experiments \citep{hansen2018a}. In the experimental set-up shown in Fig. \ref{fig:setup} the colormap on the cylinder corresponds to the laser irradiation profile. The 40 driving beams lead to a nearly uniform irradiation region close to $\unit[700]{TW~cm^{-2}}$ on the central $\sim$\unit[650]{\micro m} length portion along the target (shown in red).

The implosion dynamics were recorded with two orthogonal XRFCs - one oriented along the axial \los (view along the axis of the tube) and another along a perpendicular \los (view of the tube from the side). Each \xrfc used a 4$\times$4 pinhole array ($\unit[10]{\micro m}$ pinhole diameter) coupled with a 4-strip microchannel plate and an optical CCD, providing up to 16 images in each camera covering the whole duration of the implosion. The delay between the images within each strip was 50 ps for both lines of sight. The exposure time of each frame was \unit[200]{ps} for the axial view, whereas the perpendicular view had a \unit[50]{ps} exposure. A $\unit[635]{\micro m}$-thick Be filter was added to both XRFCs, limiting their spectral range to energies above $\sim \unit[2]{keV}$. The magnifications were $M$=2 and $M$=6 respectively. Taking this into account together with the pinhole size, instrument response and pixel size of each camera, yields a resolution of 18 and $\unit[12]{\micro m}$ for the axial and perpendicular views respectively. Other diagnostics included neutron diagnostics and X-ray emission spectroscopy of the argon dopant within the fuel. A more comprehensive study of the results from these diagnostics will be presented in future publications.

In the magnetized cases, a seed B-field of $B_0=\unit[24]{T}$ was applied along the axis of the cylinder by means of the \MIFEDS pulsed-power device \cite{Gotchev_2009}. In these cases, the axial line of sight was blocked by the \MIFEDS, and only a \xrfc perpendicular to the axis of the cylinder was used.

To model the implosions, we performed 2-dimensional extended-\MHD simulations using the Gorgon code \cite{ciardi2007, chittenden2004, walsh2017}. The specific characteristics of these simulations are given in detail in our previous publication (Walsh \textit{et al.}, 2022) \cite{walsh2022}. Our results suggest that, while the implosion dynamics are independent of the B-field before \unit[1.4]{ns}, there is a significant difference in the density of the compressed fuel, which translates in a difference in the compressed radius between the magnetized ($\sim \unit[6]{\micro m}$) and the non-magnetized ($\sim \unit[4]{\micro m}$) implosions. This is due to compression of the seed B-field, which is \textit{frozen-in} to the imploding plasma and exceeds $\unit[10]{kT}$ at stagnation. Collisional energy losses are heavily reduced in this magnetized regime, increasing the temperature in the core and hence the thermal pressure. Magnetic pressure is also significant in the magnetized implosions, increasing core pressure and reducing the overall level of compression \cite{walsh2022}.

\section{X-ray imaging data and discussion}
\label{sec:results}

An example of \xrfc data is shown in Fig. \ref{fig:XRFC_Data}, where the top half of the image shows 8 frames from the axial view and the bottom part corresponds to 8 frames from the perpendicular view. X-ray emission is observed as early as $\sim$\unit[0.39]{ns} from the axial view, where $t=0$ corresponds to the start of the laser drive. Fig. \ref{fig:XRFC_Synthetic} shows a composition of synthetic images produced by postprocessing the 2D Gorgon simulations. To do so, we applied free-free radiation transport (accounting for both the core and the shell) in either the axial or normal direction of the cylinder, and corrected for the instrument filtering and resolution. This mimics the observable data and permits a direct comparison. Hereafter, we will refer to the postprocessed results as \textit{apparent}.

Two different metrics were used to analyze these images: the separation between the two intensity peaks coming from the imploding shell and the width of the compressed core. These two metrics are not always available since core emission is negligible at early times but dominates over shell emission later in time (see Fig. \ref{fig:XRFC_Data_Image}). The two metrics therefore provide information over two different periods of time and are not directly comparable. A compilation of measured shell and core radii with (red points) and without (blue points) a seed B-field, using these two metrics is shown in Figs.~\ref{fig:XRFC_Plot}a-b respectively. The shell data were obtained from 9 shots using a combination of both \xrfc lines of sight, whereas the core measurements were obtained from 5 of the shots along the perpendicular line of sight. For the cases with a seed B-field, no data are available at early times, since the axial view was blocked by the \MIFEDS coils. The vertical error bars correspond to the resolution of the images. This translates to an error of $\sim\pm\unit[6]{\micro m}$ for the perpendicular view and $\sim\pm\unit[9]{\micro m}$ for the axial view. The horizontal error bars are related to the exposure time from each \xrfc. The shell and core radii with and without a seed B-field show no significant difference and, overall, the data is highly reproducible.

The purple dashed line in Fig. \ref{fig:XRFC_shell} corresponds to the trajectory of the shell as predicted from the \MHD simulations, whereas the green and brown lines correspond to the \textit{apparent} position of the shell for both views, once radiation transport and instrument response are taken into account. In this case, we used the same metrics as for the experimental data in order to have a direct comparison. Each of these lines is only shown for the times where data with the corresponding view were obtained. It can be seen how, although there is a $\sim\unit[50]{\micro m}$ jump in the apparent radius of the target when switching from axial to normal view, there is good agreement between experimental and synthetic data for the whole duration of the implosion, while there is systematic shift with respect to the predicted shell position (purple dashed line). 

\begin{figure}
    \centering
    \includegraphics[width=0.76\columnwidth]{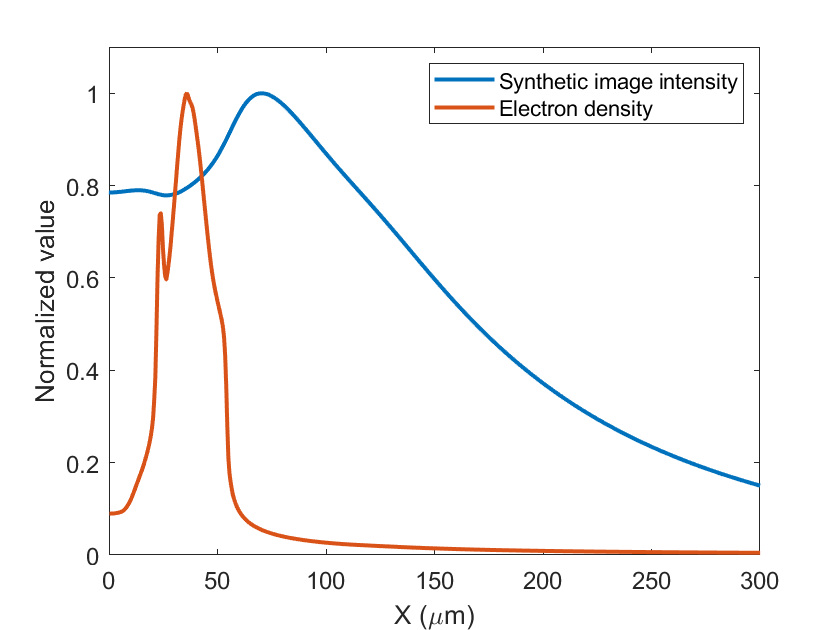}
    \caption{Normalized lineouts through the center of the target of the intensity on the synthetic postprocessed image (blue) and the electron density from the \MHD simulation (red). These correspond to $t=\unit[1.40]{ns}$ for the case with B-field=\unit[0]{T}. The densest parts of the plasma do not correspond to the brightest intensity, leading to an apparent radius on the detector further from the axis of the cylinder.}
    \label{fig:Profiles_at_1.40}
\end{figure}

This postprocessing is crucial in order to compare data and simulations. If we only considered bremsstrahlung emission, the peak signal on the detector would correspond to the densest part of the plasma. In reality, however, this is not the case, owing to radiation transport. The effect of opacity is non negligible, and therefore, the densest parts of the shell partly absorb the bremsstrahlung emission thus shifting the position of the peak intensity outwards with respect to the position of peak density. As illustration, Fig. \ref{fig:Profiles_at_1.40}, compares two lineouts through the center of the cylinder at $\unit[1.40]{ns}$. The blue line corresponds to the intensity that reaches the detector, as obtained from the postprocessed simulations, whereas the red line corresponds to the electron density profile. It can be seen how the most intense region in the detector is shifted by $\sim\unit[30]{\micro m}$ with respect to the position of the shell (peak of the electron density profile).

A particularly interesting consequence of the combined effect of radiation transport and the instrument response is that, at early times during the implosion ($\sim \unit[0.4]{ns}$) the apparent radius of the shell is larger than the initial target radius ($\unit[300]{\micro m}$). This is observed in both the experimental and the synthetic data with an axial view (green line). This fact is not a direct consequence of radiation transport, as it only appears when the Be filtering in the detector is taken into account. The mid-energy emission from the dense plasma is heavily absorbed by this filter, prior to reaching the detector, whereas the high energy photons ($\unit[5-20]{keV}$) from the ablated plasma reach the detector without being absorbed neither by the plasma nor by the Be filtering.

The black line in Fig. \ref{fig:XRFC_shell} shows the mean trajectory from all the data points. Linear behaviour observed between $\sim0.4$ and $\sim\unit[1.1]{ns}$ was used to estimate an apparent implosion velocity of $\unit[200\pm10]{km/s}$ at early times. The implosion then accelerates from $\sim \unit[1.1]{ns}$ onward, reaching a velocity of $\unit[280\pm40]{km/s}$. These values are consistent with the apparent velocity from the postprocessed simulations, and the results from previous work \cite{hansen2018a}.

Figure \ref{fig:XRFC_core} shows measurements of the compressed cylinder radius. The convergence ratio ($CR=R/R_0$) was estimated by taking the mean core radii of $R\sim\unit[15]{\micro m}$, resulting in a value of $CR\sim 20$. The solid lines in this figure correspond to the values predicted by Gorgon for both unmagnetized (blue) and magnetized (red) implosions up to stagnation ($t=$\unit[1.50]{ns}). Given that the measured core radius is also limited by the spatial resolution of the \xrfc and the blurring effect of radiation transport, the simulated values are not directly comparable with the experimental data. In order to establish a direct comparison, the dotted lines correspond to the core radius obtained from the postprocessed simulation (using the same color code). Note that, once these effects are taken into account, the differences in compression owed to the B-field are indiscernible and the lines overlap.

We observe a difference between experiment and simulations, where the experimentally-observed compressed radius is $\sim 1.9\times$ larger than the predictions from extended \MHD simulations. This discrepancy may have several causes which are currently being investigated, including target pre-heat caused by hot electrons, mixing of the ablator into the fuel, limitations of 2D versus 3D modelling, or azimuthal hydrodynamic instabilities that lie below the \xrfc resolution - all of which may contribute towards reduced implosion performance. A similar convergence discrepancy between 2D \MHD simulations and experiments has been previously reported in analogous cylindrical implosions at OMEGA~\cite{davies2019}, with predicted areal densities $2-3\times$ higher than measurements.

\section{Conclusions}
\label{section:Conclusions}

We have measured the radial compression of \unit[14.5]{kJ} laser-driven cylindrical implosions with and without an applied B-field of \unit[24]{T}, combining x-ray images taken from axial and normal views. Our results indicate that the implosion speeds up at $\sim\unit[1.1]{ns}$, accelerating from \unit[200]{km/s} to \unit[280]{km/s}. Stagnation occurs at $\sim\unit[1.5]{ns}$ and lasts for $\sim\unit[200]{ps}$. 

The X-ray imaging data agree with extended \MHD simulations once radiation transport and instrument filtering are taken into account. We find that these effects are crucial to the interpretation of the data, since they result in an apparent size of the cylinder which is larger than in reality.

Additionally, a significant difference was observed in the compressed core radius at stagnation, with the postprocessed \MHD simulations predicting a smaller apparent radius than observed.

\section*{Acknowledgements}

This work was performed under the auspices of the U.S. Department of Energy by Lawrence Livermore National Laboratory under Contract DE-AC52-07NA27344. This document was prepared as an account of work sponsored by an agency of the United States government. Neither the United States government nor Lawrence Livermore National Security, LLC, nor any of their employees makes any warranty, expressed or implied, or assumes any legal liability or responsibility for the accuracy, completeness, or usefulness of any information, apparatus, product, or process disclosed, or represents that its use would not infringe privately owned rights. Reference herein to any specific commercial product, process, or service by trade name, trademark, manufacturer, or otherwise does not necessarily constitute or imply its endorsement, recommendation, or favoring by the United States government or Lawrence Livermore National Security, LLC. The views and opinions of authors expressed herein do not necessarily state or reflect those of the United States government or Lawrence Livermore National Security, LLC, and shall not be used for advertising or product endorsement purposes. 

This work has been carried out within the framework of the EUROfusion Consortium, funded by the European Union via the Euratom Research and Training Programme (Grant Agreements No. 633053 and No. 101052200 — EUROfusion). Views and opinions expressed are however those of the author(s) only and do not necessarily reflect those of the European Union or the European Commission. Neither the European Union nor the European Commission can be held responsible for them. The involved teams have operated within the framework of the Enabling Research Projects: AWP17-ENR-IFE-CEA-02 and AWP21-ENR-IFE.01.CEA.

This material is based upon work supported by the US National Nuclear Security Administration and National Laser Users' Facility under Award No. DE-NA0003940, and by the US Department of Energy - Office of Science under Grant No. DE-SC0022250. The work has also been supported by the Research Grant No. CEI2020-FEI02 from the Consejería de Economía, Industria, Comercio y Conocimiento del Gobierno de Canarias; and by Research Grant No. PID2019-108764RB-I00 from the Spanish Ministry of Science and Innovation.

This study has received financial support from the French State in the framework of the Investments for the Future programme IdEx université de Bordeaux / GPR LIGHT. F.S.-V. acknowledges funding from The Royal Society (UK) through a University Research Fellowship. C.V. acknowledges the support from the LIGHT S\&T Graduate Program (PIA3 Investment for the Future Program, ANR-17-EURE-0027). 

\noindent \textbf{Data availability:} The data presented in this paper may be obtained from the authors upon reasonable request. \textbf{Conflicts of interest:} The authors declare no conflicts of interest.


\bibliography{sample}

\end{document}